\documentclass[pra,aps,twocolumn]{revtex4}
\usepackage{xr}
\usepackage{graphicx,amsmath}
\usepackage{txfonts}
\usepackage[colorlinks, citecolor=blue,linkcolor=blue]{hyperref}

\begin{document}
\date{\today}

 \title{Feedback Control of Two-mode Output Entanglement and Steering in Cavity Optomechanics }
\author{Muhammad Asjad, Paolo Tombesi and David Vitali}
 \affiliation{School of Science and Technology, Physics Division, University of Camerino, 62032 Camerino (MC), Italy, and INFN, Sezione di Perugia, Perugia, Italy}
\begin{abstract}

We show that the closed-loop control obtained by feeding back the derivative of the signal from the homodyne measurement of one mode of the light exiting a two-mode optical cavity interacting with a mechanical resonator permits to control and increase optical output entanglement. In particular, the proposed feedback-enhanced setup allows to achieve a fidelity of coherent state teleportation greater than the threshold value of 2/3 for secure teleportation, and two-way steering between the two cavity's output modes down the line in presence of loss, which otherwise would not be possible without feedback.
\end{abstract}
\maketitle
\section{Introduction}
Generally to control a system both in classical and quantum mechanics one uses either closed-loop or open-loop control. Here we will focus on quantum mechanical closed-loop control with classical feedback. Indeed, as well specified in Ref. \cite{Lloyd2000}, one has to distinguish between quantum control with classical feedback and fully quantum control, where the quantum feedback  controller acquires and processes quantum information. The first is also called measurement-based feedback because to control the system one has first to measure an observable and then, by means of an actuator, to feed the system with the result of the measurement, eventually manipulated in some way, i.e. using classical information. Refs. \cite{Wiseman1993a, Wiseman1993} first described how to use feedback in all-optical cases. After these pioneering works in optical domain several other proposals were introduced to show how with feedback one could simulate the presence of a squeezed environment \cite{Tombesi1994}, slow down the destruction of macroscopic coherence \cite{Tombesi1995}, and be useful to control the environment's thermal fluctuations so to cool a mechanical oscillator \cite{Mancini1998,Courty2001,Vitali2002,Vitali2003,Genes2009}. This last proposal has been implemented in many optomechanical systems~\cite{Cohadon1999,Metzger2004,Kleckner2006,Arcizet2006b,Poggio2007,Corbitt2007,Corbitt2007a,
 Wilson2015,Krause2015,Sudhir2016}, and on a trapped ion~\cite{Bushev2006}. Moreover, parametric feedback schemes have been proposed and implemented for cooling and trapping single atoms~\cite{Koch2010} and trapped nanospheres~\cite{Gieseler2012,Genoni2015}.
In the case of superconducting qubits, feedback schemes based on parity measurements have been recently demonstrated~\cite{Riste2013} achieving deterministic generation of entanglement.

In this paper we will show that one can use feedback to control and improve continuous variable (CV) entanglement: by adjusting the feedback gain one can enhance the value of CV entanglement, measured through the logarithmic negativity \cite{Vidal2002,Eisert2001, Adesso2004,Plenio2005}.
As an example of the utility of our result we show that the fidelity of CV teleportation can be enhanced and surpass the threshold value for quantum teleportation $F_{thr}=2/3$ \cite{Grosshans2001} even at the end of a lossy channel, just by implementing the appropriate feedback control. We also show how to realize a two-way steerable Gaussian bipartite state, so that the sent information is really secure  \cite{He2015}, even though a cheating sender has cloned with an optimal cloning machine \cite{Buzek1996} the state to be teleported and has given the cloned state to an eavesdropper.

The paper is organized as follows: In Sec.~\ref{MD} we describe the model by introducing the quantum Heisenberg-Langevin equations. In Sec.~\ref{afb} we add the feedback and discuss the basic dynamics of the system in the presence of feedback. In Sec.~\ref{afb} we derive the explicit expression for the covariance matrix of the filtered output cavity modes in the presence of feedback force. In Sec.~\ref{ses} we report the numerical results of steady state entanglement between two filtered output optical modes and two way steerability in the presence of feedback. In Sec.~\ref{fd3} we consider a different way of adding the feedback, by introducing a third optical mode and use it for feedback in order to control the entanglement between the other two filtered output optical modes. Finally, in Sec.~\ref{con} we compare the two different feedback schemes and draw some conclusions.

\section{Model and Discussion}\label{MD}
\begin{figure}[tb]
 \centering
\includegraphics[width=.45\textwidth]{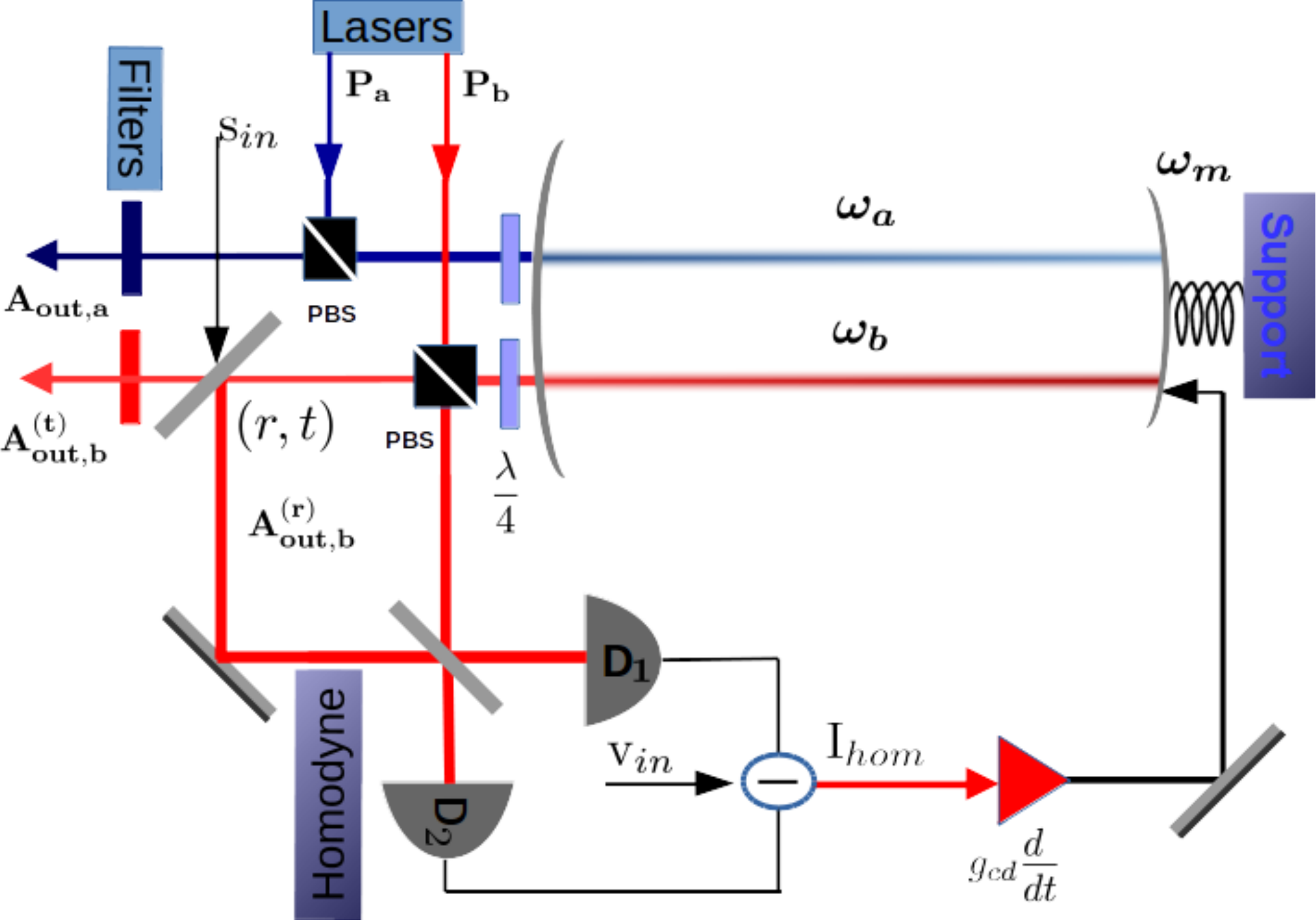}
\caption{(Color online) Description of the homodyne-based feedback scheme} \label{scheme}
\end{figure}

We consider the multipartite optomechanical set-up shown in Fig.\ref{scheme}. A bichromatic field at two different frequencies $\omega_{Lj}/2\pi$ $(j=a,b)$ with powers $P_j$ drives two cavity modes of frequencies $\omega_j/2\pi$ both interacting with a mechanical resonator oscillating at frequency $\omega_m/2\pi$.
One of these optical modes is homodyned and used for feedback, then the two output modes are filtered and form the two fields of interest for quantum communication, and the mechanical mode mediates the necessary interaction between the optical modes. The entangled output fields can be used down a lossy channel for teleportation and we show that feedback helps in obtaining a fidelity higher than the security threshold, which would have not been achievable without feedback with the same system's parameter values.
The full Hamiltonian of the optomechanical system composed by two optical modes, Mode $A$, Mode $B$, and one mechanical oscillator with effective mass $m$, considered in the frame rotating for each optical mode at the corresponding driving laser frequency $\omega_{Lj}/2\pi$, is given by \cite{Genes2009,Giovannetti2001a},
\begin{equation}
\hat{H}=\dfrac{\hbar\omega_m}{2}(\hat{p}^2 +\hat{q}^2)+\hbar\Sigma_j[(\delta_j
- g_j \hat{q})\hat{A}^\dagger_j \hat{A}_j
+i(E_{j}\hat{A}^\dagger_j-H.c.)],
\label{hml}
\end{equation}
where $\delta_j=\omega_{j}-\omega_{Lj}$ is the detuning of the laser frequency from the cavity frequency $\omega_j/2\pi$, the dimensionless mechanical resonator position $\hat{q}$ and momentum $\hat{p}$ have commutation relation $[\hat{q},\hat{p}]=i$, $\hat{A}_j (\hat{A}^\dagger_j)$ is the annihilation (creation) operator for the cavity mode $j$ with commutation relation $[\hat{A}_j,\hat{A}^\dagger_k]=\delta_{jk}$, and $g_j= x_{zpf}\sqrt{2}(d\omega_j/dx)$ is the bare single photon optomechanical coupling, where $x_{zpf}= \sqrt{\hbar /2 m\omega_m}$ represents the zero-point position fluctuations of the mechanical oscillator. $E_j$ is the amplitude of the $j$-th driving field, and we will deal with only a single mechanical mode \cite{Pinard1999}.

In order to study the full dynamics of the system we use the Heisenberg-Langevin equations of motion~\cite{Gardiner2000}, adding the effect of damping and noise to the evolution driven by the Hamiltonian of Eq.~(\ref{hml}), obtaining
\begin{subequations}\label{qles}
\begin{eqnarray}
\dot{\hat q}&=&\omega_m \hat  p,\\
\dot{\hat p}&=&-\omega_m \hat q-\gamma_m \hat p+\sum_{i} g_i \hat{A} ^\dagger_i \hat{A} _i+\zeta(t),\\
\dot{\hat A}_a&=&-(\kappa_a+i\delta_a) \hat{A} _a+ig_a \hat{A} _a \hat{q} +E_a+ \sqrt{2\kappa_a}\hat{A} ^{in}_a,\\
\dot{\hat A}_b&=&-(\kappa_b+i\delta_b) \hat{A} _b+ig_b \hat{A} _b \hat{q} +E_b+ \sqrt{2\kappa_b}\hat{A} ^{in}_b,
\end{eqnarray}
\end{subequations}
where, $\gamma_m$ is the damping rate of the mechanical oscillator and  $\zeta(t)$ is the thermal fluctuating force associated with the mechanical damping and characterized by the correlation function $(\langle\zeta(t)\zeta(t')\rangle+\langle\zeta(t')\zeta(t)\rangle)/2=\gamma_m(2n_{th}+1)\delta(t-t')$, where $n_{th}=(e^{(\hbar\omega_m/K_b T)}-1)^{-1}$ is the equilibrium thermal occupation number of the mechanical resonator.
We also introduce the optical input noises of Mode A and Mode B, respectively given by  ${\hat A} ^{in}_a$ and ${\hat A} ^{in}_b$, with the following correlation functions
\begin{subequations}\label{corr}
\begin{eqnarray}
\langle {\hat A} ^{in}_a(t) {\hat A} ^{in^\dagger }_a(t') \rangle &=& (N(\omega_a)+1)\delta(t-t'), \\
\langle {\hat A} ^{in^\dagger }_a(t) {\hat A} ^{in}_a(t') \rangle &=& (N(\omega_a))\delta(t-t'), \\
\langle {\hat A} ^{in}_b(t) {\hat A} ^{in^\dagger }_b(t') \rangle &=& (N(\omega_b)+1)\delta(t-t'), \\
\langle {\hat A} ^{in^\dagger }_b(t) {\hat A} ^{in}_b(t') \rangle &=& (N(\omega_b))\delta(t-t'),
\end{eqnarray}
\end{subequations}
where $N(\omega_j)=[e^{(\hbar\omega_j/k_b T)}-1]^{-1}$ with $j= a, b$ is the mean thermal photon number. At optical frequencies $(\hbar\omega_j/k_b T \gg 1)$, therefore, one can safely assume $N(\omega_j)\approx 0$.
We consider the regime where both Mode A and Mode B are strongly driven and the field inside the cavity is very intense. In this case the dynamics of the system can be described by quantum fluctuations around the steady state, which is stable with the right choice of the various system's parameters. Therefore, one can make the semiclassical approximation to linearize the system of nonlinear Langevin equations by writing each operator of the system as the sum of its steady state value and a small fluctuation, $\hat{ A} _j=A_{j s}+\delta \hat{ A} _{j}$,  $ \hat{q}=q_{s}+\delta  \hat{q}$, $\hat{p}=p_{s}+\delta  \hat{p}$. The parameters $A_{j s}$, $p_s$  and $q_s$ are the solutions of the nonlinear algebraic equations obtained by factorizing Eqs.~(\ref{qles}) and setting the time derivatives equal to zero: $p_{s}=0$,
\begin{eqnarray}
q_{s}= \sum_{j=a,b}\dfrac{g_j |A_{js}|^2}{\omega_m},\qquad
A_{js}= \dfrac{E_j}{\kappa_j+i\Delta_j},
\end{eqnarray}
where $\Delta_j=\delta_j-\sum g_j q_s$ is the effective detuning. Then by defining the quadrature fluctuations of the cavity field $\delta  \hat{X}_j=(\delta  \hat{A}_j+\delta   \hat{A}^\dagger_j)/2$ and $\delta \hat{Y}_i=(\delta  \hat{A}_j-\delta \hat{A}^\dagger_j)/i\sqrt{2}$, and the corresponding input noise quadratures $ \hat{X}^{in}_j=( \hat{A}^{in}_j+ \hat{A}^{\dagger\, in }_{j})/\sqrt{2}$ and $ \hat{Y}^{in}_j=( \hat{A}^{in}_j- \hat{A}^{\dagger\, in }_{j})/i\sqrt{2}$, the linearized quantum Heisenberg-Langevin equations in compact form are given by
\begin{equation}  \label{sys1}
\dot{\hat {\bf {R}}}(t)={\bf {A}}^{dr} {\hat{\bf {R}}}(t)+{\hat{\bf {n}}}(t)
\end{equation}
where $\hat{\bf {R}}(t)=[\delta  \hat{q},\delta \hat{p}, \delta  \hat{X}_a,\delta  \hat{Y}_a, \delta  \hat{X}_b,\delta  \hat{Y}_b]^\mathsf{T}$ (where $\mathsf{T}$ denotes transposition)
 is the vector of the system's fluctuations quadratures, ${\hat{\bf{n}}}(t)=[0,\zeta(t),\sqrt{2\kappa_a}  \hat{X}^{in}_a,\sqrt{2\kappa_a}  \hat{Y}^{in}_a,\sqrt{2\kappa_b}\hat{X}^{in}_b,\sqrt{2\kappa_b}\hat{Y}^{in}_b]^\mathsf{T}$ is the corresponding vector of noises and ${\bf A^{dr}} $ is the drift matrix, which is given by
\begin{equation}
{\bf A^{dr}}=\left( \begin{array}{cccccc}
0& \omega_m& 0& 0& 0& 0\\
-\omega_m& -\gamma_m& G_a& 0& G_b&0\\
0& 0& -\kappa_a& \Delta_a& 0& 0 \\
G_a& 0& -\Delta_a& -\kappa_a& 0& 0\\
0& 0& 0& 0& -\kappa_b& \Delta_b\\
G_b& 0& 0& 0& -\Delta_b& -\kappa_b \\
\end{array}\right). \label{val}
\end{equation}
where $G_j=g_j |A_{js}|\sqrt{2}$ is the dressed optomechanical coupling for  $j$-th mode.

In the case of a tripartite optomechanical system, various proposals already showed that the two output modes can be strongly entangled, e.g., two orthogonal modes with the same frequency~\cite{Giovannetti2001a}, two optical modes at different frequencies~\cite{Genes2009, Wang2012,Tian2012,Kuzyk2013,Asjad2015a}, or an optical and microwave mode~\cite{ Barzanjeh2011, Barzanjeh2012}; in Refs.~\cite{Asjad2015a, Barzanjeh2011, Barzanjeh2012} entanglement was sufficiently good that it could be exploited to perform CV teleportation.
We now show that the value of CV entanglement can be controlled and enhanced via feedback, and to this end we use a fraction of one of the optical modes and homodyne it.

\section{Adding feedback}\label{afb}

We consider cold damping feedback~\cite{Mancini1998,Courty2001,Vitali2002,Genes2009,Cohadon1999,Metzger2004,Kleckner2006,
Arcizet2006b,Poggio2007,Corbitt2007,Corbitt2007a,Poggio2007, Wilson2015,Krause2015,Sudhir2016} in which the position of the oscillator is measured by means of a phase-sensitive detection of the output of cavity Mode B. For this purpose a beam splitter is used which splits the output of Mode B into a transmitted and reflected field. The transmitted part may be used by a generic quantum communication protocol, while the reflected part is used to measure the position of the oscillator by means of a phase-sensitive detection, and is then fed back to the oscillator by applying a force whose intensity is proportional to the time derivative of the output signal, that is, to the oscillator velocity (the cold damping technique \cite{Mancini1998,Courty2001,Vitali2002,Genes2009,Cohadon1999,Metzger2004,Kleckner2006,Arcizet2006b,Poggio2007,Corbitt2007,Corbitt2007a,Poggio2007, Wilson2015,Krause2015,Sudhir2016}). Real time monitoring of the resonator position is provided by the homodyne measurement of the phase quadrature $\delta\hat {Y}^{(r)}_{b}(t)=r \delta \hat {Y}_b^{out}(t)+t \hat {Y}^{in}_{s}(t)$, where $\delta \hat{Y}_b^{out}$ is the phase quadrature of the output field fluctuation, which is obtained by using the input-output relation $\delta \hat{Y}_b^{out}= \sqrt{2\kappa_b} \delta   \hat{Y}_b- \hat{Y}_b^{in}$ \cite{Gardiner2000}, and $ \hat {Y}^{in}_{s}(t)=(\hat{s}_{in}(t)- \hat {s}^\dagger_{in}(t))/\sqrt{2}i$, with $ \hat {s}_{in}(t)$ the vacuum noise entering the unused input port of the beam splitter (with the usual commutation relation $[ \hat {s}_{in}(t), \hat {s}^\dagger_{in}(t')]=\delta(t-t')$). Moreover, $r$ and $t$ are the reflection and transmission coefficient of the beam splitter, respectively, with $r^2+t^2=1$. The feedback loop is described by an additional force term on the equation of motion of the mechanical momentum $\{\delta\dot{\hat{p}}(t)\}_{fb}$ given by
\begin{eqnarray}\label{fbe}
\{\delta\dot {\hat{p}}(t)\}_{fb}&=&\dfrac{i}{\sqrt{2\kappa_b}}\dfrac{d}{dt}(\delta\hat {Y}^{(hom)}_{b}(t))[g_{cd}\delta \hat{q},\delta \hat{p}]\nonumber\\
&=&-\dfrac{g_{cd}}{\sqrt{2\kappa_b}}\dfrac{d}{dt}\delta\hat {Y}^{(hom)}_{b}(t),
\end{eqnarray}
where $g_{cd}>0$ is the feedback gain and $\delta\hat {Y}^{(hom)}_{b}(t)$ the detected field quadrature. If the detector efficiency is $\sigma$ then the detected field quadrature can be represented by the operator
\begin{eqnarray} \label{hom}
\delta\hat {Y}^{(hom)}_{b}(t)&=&\sqrt{\sigma} \delta \hat {Y}^{(r)}_{b}(t)+\sqrt{1-\sigma} \hat {Y}^{in}_{v}(t),
\end{eqnarray}
where $ \hat {Y}^{in}_{v}(t)=(\hat{v}_{in}(t)- \hat {v}^\dagger_{in}(t))/\sqrt{2}i$ is the Gaussian noise operator associated with the non-unit homodyne detection efficiency, with correlation $\langle \hat{Y}^{in}_{v}(t) \hat{Y}^{in}_{v}(t')\rangle=\delta(t-t')$.
Inserting Eq. (\ref{hom}) into Eq. (\ref{fbe}) and then adding the resulting feedback force into the linearized quantum Langevin equations of
Eqs.~(\ref{sys1}), the dynamics of the three-mode optomechanical system modified by the feedback force can be written in the frequency domain in the following compact matrix form
\begin{eqnarray}\label{fb}
{\bf \hat{R}}^{fb}(\omega)= -{\bf M}(\omega){\bf \hat{N}}^{fb}(\omega),
\end{eqnarray}
where ${\hat{\bf R}}^{fb}(\omega)=[\delta \hat{q}^{fb},\delta  \hat{p}^{fb}, \delta  \hat{X}^{fb}_a,\delta  \hat{Y}^{fb}_a, \delta \hat{ X}^{fb}_b,\delta  \hat{Y}^{fb}_b]^\mathsf{T}$ is the Fourier transform of the vector with CV internal quadrature fluctuations in the presence of feedback, ${{\bf \hat{N}}}^{fb}(\omega)={\hat{\bf{n}}}(\omega)+{\hat{\bf{n}}}_{fb}(\omega)$ is the corresponding vector of input noises in presence of feedback with ${\hat{\bf{n}}}_{fb}(\omega)=[0,-\sqrt{\sigma}r g_{cd}\sqrt{2\kappa_b}\,\hat{Y}^{in}_b(\omega)(1+ \dfrac{i\omega}{2\kappa_b})+\dfrac{i\omega \sqrt{\sigma}tg_{cd}}{\sqrt{2\kappa_b}}\,\hat{Y}^{in}_{s}(\omega)+\dfrac{i\omega \sqrt{1-\sigma}g_{cd}}{\sqrt{2\kappa_b}}\, \hat{Y}^{in}_{v}(\omega),0,0,0,0]^\mathsf{T}$ and ${\bf M}(\omega)=(i\omega{\bf I}+{\bf{A}}^{fb})^{-1}$ with $\bf I$ the 6x6 identity matrix and ${\bf{A}}^{fb}$ the drift matrix of the linearized dynamical system modified by feedback. This latter matrix can be written as ${\bf A}^{fb}= {\bf A}^{dr}+ {\bf F}^{dr}$, where ${\bf F}^{dr}$ is a $6\times6$ matrix whose non zero elements are $\{{\bf F}^{dr}\}_{(2,2)}=-G_{cd }G_b$,  $\{{\bf F}^{dr}\}_{(2,5)}=G_{cd }\Delta_b $ and $\{{\bf F}^{dr}\}_{(2,6)}=G_{cd }\kappa_b $,
%
with $G_{cd }=\sqrt{\sigma}rg_{cd}$.

The tripartite optomechanical system is stable and reaches its steady state only if all the eigenvalues of the drift matrix ${\bf {A}}^{fb}$ have negative real part. The stability conditions, which now depend on the feedback gain, can be obtained by applying the Routh-Hurwitz criterion \cite{Gopal2002}, which however is too cumbersome to be explicitly reported here. In particular, one has to verify that the effective mechanical damping constant remains positive due to the combined and opposite action of the feedback force and of the backaction of mode $B$. We always consider steady state entanglement and therefore we always consider a parameter regime where the system is stable.

\begin{figure*}[t!]
\centering
\includegraphics[width=1\textwidth]{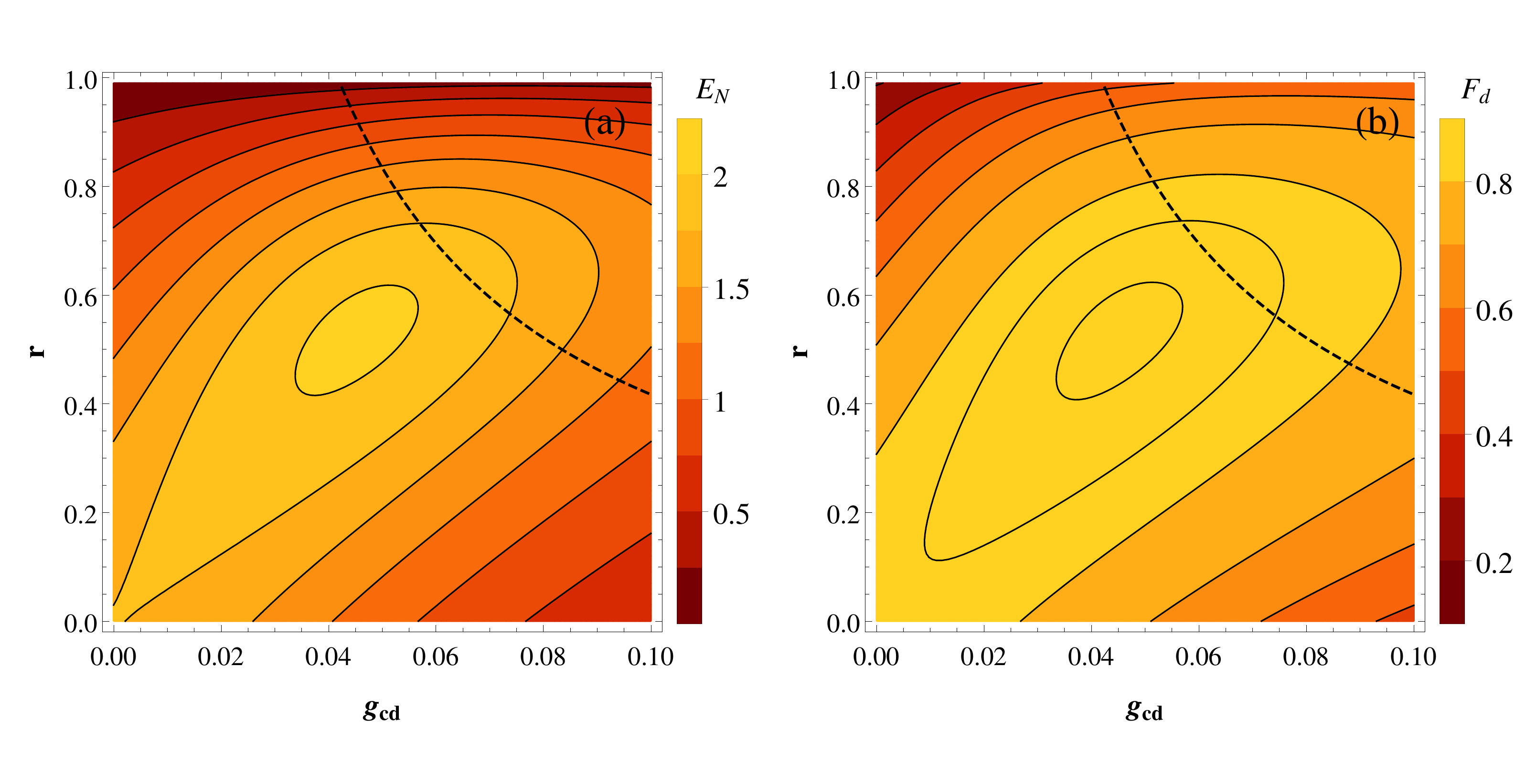}\\
\caption{(Color online) Logarithmic negativity $E_N$ (a), and fidelity of teleportation of a coherent state (b), as a function of reflection coefficient of beam splitter $r$ $(t=\sqrt{1-r^2})$ and feedback gain $g_{cd}$. The optimal values for $r$ and $g_{cd}$ are essentially the same for entanglement and fidelity. The dashed curve corresponds just to the condition $g_{cd}~r=G_b/(\sqrt{\sigma}\omega_m)$, when cold damping feedback cancels the heating effect of the blue-detuned mode B. The system parameters are $\omega_m=2\pi\times10\,\mathrm{MHz}$, $\gamma_m=1.5\times10^{-5}\,\omega_m$,  $\kappa_a=0.01\,\omega_m,\kappa_b=0.01\,\omega_m$, temperature $T=400\,mK$, $\omega_m \tau_b=\omega_m \tau_a=2000$, $G_a=0.065\,\omega_m$, $G_b=0.04\,\omega_m$, $\Delta_a=\omega_m$, $\Delta_b=-\omega_m$, $\Omega_{a}=-\Omega_{b}=\omega_m$ and $\sigma=0.92$.
} \label{ab_r}
\end{figure*}
\subsection{Covariance Matrix of the Filtered Output Quadratures in the presence of Feedback}\label{Cov}

We want to study the entanglement of the traveling optical mode $A$ fluctuations and the transmitted fluctuations of the mode $B$ at the output of the optomechanical cavity.
By using the input-output relation, the spectral components of the output field's quadratures in the presence of the feedback force are given by
\begin{eqnarray}\label{oR}
{\hat{\bf R}}^{out}(\omega)&=& {\bf T}_t({\bf P} {\hat{\bf R}}^{fb}(\omega)-{\hat{\bf N}}(\omega))-{\bf T}_r {\bf \hat{N}}_{s}(\omega)\nonumber \\
&=&{\bf T}_t\left(-{\bf P} {\bf M}(\omega){\hat{\bf N}}^{fb}(\omega)-{\hat{\bf N}}(\omega)\right)-{\bf T}_r {\bf \hat{N}}_{s}(\omega),\nonumber \\
\end{eqnarray}
 where ${\hat{\bf R}}^{out}(\omega)=[\delta \hat{q}^{fb},\delta  \hat{p}^{fb}, \delta  \hat{X}^{out}_a,\delta  \hat{Y}^{out}_a, \delta \hat{ X}^{(out)}_b,\delta  \hat{Y}^{(out)}_b]^\mathsf{T}$,
${\bf P}= Diag[1,1,\sqrt{2\kappa_a},\sqrt{2\kappa_a},\sqrt{2\kappa_b},\sqrt{2\kappa_b}]$,
${\hat{\bf N}}(\omega)=[0,0,\hat{X}^{in}_a(\omega),\hat{Y}^{in}_a(\omega),\hat{X}^{in}_b(\omega),\hat{Y}^{in}_b(\omega)]^\mathsf{T}$,
 ${\bf T}_t=Diag[1,1,1,1,t,t]^\mathsf{T}$, ${\bf T}_r=Diag[1,1,1,1,r,r]^\mathsf{T}$ and ${\bf\hat{ N}}_{s}(\omega)=[0,0,0,0, \hat{X}^{in}_{s}(\omega), \hat{Y}^{in}_{s}(\omega)]^\mathsf{T}$.

As shown in Ref.~\cite{Genes2009}, the correlation between the output optical modes can be optimized with filters. The field's filtered mode can be defined as
\begin{equation}
\delta  \hat{A}^{flt}_j(t)= \int^t_{-\infty} h_j(t-t')\delta  \hat{A}^{out}_j(t') dt' \qquad (j=a,b)
\end{equation}
 where $\delta  \hat{A}^{flt}_j(t)$ is the corresponding bosonic annihilation operator at the output of the j-th causal filter $h_j(t)$. The explicit form of the $h_j(t)$ in time and frequency domain can be written as \cite{Genes2009}
\begin{equation}\label{filt}
h_j(t)=\dfrac{e^{-(1/\tau_j+i\Omega_j)t}}{\sqrt{2/\tau_j}}\theta_j(t)\,\,\, \text{and}\,\,\, h_j(\omega)=\dfrac{\sqrt{\tau_j/\pi}}{1+i\tau_j(\Omega_j-\omega)}
\end{equation}
where $\tau_j$ and $\Omega_j$ are the inverse bandwidth and central frequency of the j-th filter. $\theta_j(t)$ is the Heaviside step function. Since the steady state of the system is a zero mean Gaussian state, it is fully described by its second order correlations. The covariance matrix of the filtered output fluctuation quadratures can be written as
\begin{equation}\label{cor}
2 {\bf V}^{flt}(\omega, \omega')=\langle {\hat{\bf R}}^{flt}(\omega)\cdot {\hat{\bf R}}^{flt\mathsf{T}}(\omega')+{\hat{\bf R}}^{flt}(\omega')\cdot {\hat{\bf R}}^{flt\mathsf{T}}(\omega)\rangle.
\end{equation}
where ${\hat{\bf R}}^{flt}(\omega)= {\bf T}(\omega) {\hat{\bf R}}^{out}(\omega)$, with ${\bf T}(\omega)$ the Fourier transform of the ${\bf T}(t)$ matrix containing the filter functions, given by
\begin{equation}
{\bf T}(t)=\left(\small \begin{array}{cccccc}
\delta(t)  &0  &0  &0   &0      &0      \\
0   &\delta(t)  &0   &0  &0  &0    \\
0   &0   &Re[h_a(t)] &- Im[h_a(t)] &0      &0  \\
0   &0        & Im[h_a(t)] & Re[h_a(t)]   &0  &0  \\
0    &0   &0   &0       &Re[h_b(t)]  &-Im[h_b(t)] \\
0     &0        &0   &0       & Im[h_b(t)]  & Re[h_b(t)] \\
\end{array}
\right).
\end{equation}
By substituting ${\hat{\bf R}}^{flt}(\omega)$ in Eq.(\ref{cor}) we get
\begin{eqnarray}\label{vfl}
2 {\bf V}^{flt}(\omega, \omega')&=& {\bf T}(\omega)\langle {\hat{\bf R}}^{out}(\omega) {\hat{\bf R}}^{out\mathsf{T}}(\omega')\rangle  {\bf T}^{\mathsf{T}}(\omega')\nonumber\\
&+& {\bf T}(\omega')\langle {\hat{\bf R}}^{out}(\omega') {\hat{\bf R}}^{out\mathsf{T}}(\omega)\rangle {\bf T}^{\mathsf{T}}(\omega),
\end{eqnarray}
and the explicit form of the two frequency autocorrelation $\langle {\hat{\bf R}}^{out}(\omega) {\hat{\bf R}}^{out\mathsf{T}}(\omega')\rangle $ is given by
\begin{eqnarray} \label{coro}
\langle {\hat{\bf R}}^{out}(\omega) {\hat{\bf R}}^{out\mathsf{T}}(\omega')\rangle &=&\{{\bf T}_t[({\bf P} {\bf M}(\omega){\bf D}^{fb}(\omega,\omega'){\bf M}^\mathsf{T}(\omega'){\bf P}+ {\bf D}_1\nonumber\\
&+&{\bf P} {\bf M}(\omega) {\bf D}_2 + {\bf D}_2 {\bf M}^\mathsf{T}(\omega') {\bf P} ]{\bf T}^\mathsf{T}_t+{\bf T}_r {\bf D}_3 {\bf T}^\mathsf{T}_r\nonumber\\
&-&{\bf T}(\omega){\bf F}(\omega){\bf T}^\mathsf{T}(\omega')\}\delta(\omega+\omega'),
\end{eqnarray}
where, ${\bf F}(\omega,\omega')={\bf T_t}{\bf P} {\bf M}(\omega) {\bf D_4}(\omega) {\bf T_r}^\mathsf{T}$. Moreover, in the above equation we have defined
\begin{eqnarray}\label{nn1}
\langle {\hat{\bf N}}^{fd}(\omega){\hat{\bf N}}^{fd\mathsf{T}}(\omega')\rangle&=&{\bf D}^{fb}(\omega,\omega') \delta(\omega+\omega'),\nonumber\\
\langle {\hat{\bf N}}(\omega) {\hat{\bf N}}^\mathsf{T}(\omega') \rangle&=&{\bf D}_1 \delta(\omega+\omega'),\nonumber\\
\langle {\hat{\bf N}}^{fd}(\omega){\hat{\bf N}}^\mathsf{T}(\omega')\rangle&=&\langle {\hat{\bf N}}^\mathsf{T}(\omega) {\hat{\bf N}}^{fd\mathsf{T}}(\omega')\rangle= {\bf D}_2 \delta(\omega+\omega'),\nonumber\\
\langle {\hat{\bf N}}_{s}(\omega){\hat{\bf N}}_{s}^\mathsf{T}(\omega')\rangle&=&{\bf D_3}\delta(\omega+\omega')\nonumber\\
\langle{\hat{\bf N}}^{fd}(\omega){\hat{\bf N}}_{s}^\mathsf{T}(\omega')\rangle&=&{\bf D_4}(\omega,\omega')\delta(\omega+\omega'),
\end{eqnarray}
where, ${\bf D}^{fb}(\omega,\omega')= {\bf d}+{\bf d}^{fb}(\omega,\omega')$ is the diffusion matrix in the presence of feedback with
$${\bf d}=Diag[0,\gamma_m(2n_{th}+1),\kappa_a,\kappa_a,\kappa_b,\kappa_b]$$
%
and ${\bf d}^{fb}(\omega,\omega')$ is a $6\times6$ matrix with only three non-zero elements, given by
\begin{eqnarray}
\{{\bf d}^{fb}(\omega,\omega')\}_{22}&=&g^2_{cd} \sigma r^2 \kappa_b\left(1+\frac{i\omega}{2\kappa_b}\right)\left(1+\frac{i\omega'}{2\kappa_b}\right)\nonumber\\
&-&\frac{\omega \omega'g^2_{cd}}{4\kappa_b} \left(1-r^2\sigma \right),\nonumber\\
\{{\bf d}^{fb}(\omega,\omega')\}_{26}&=&-g_{cd}\kappa_b\left(1+ \frac{i\omega}{2\kappa_b}\right),\nonumber\\
\{{\bf d}^{fb}(\omega,\omega')\}_{62}&=&-g_{cd}\kappa_b\left(1+ \frac{i\omega'}{2\kappa_b}\right).
\end{eqnarray}
Finally ${\bf D_1}=1/2\,\, Diag[0,0,1,1,1,1]$,\, ${\bf D_2}=1/\sqrt{2}\,\, Diag[0,0,\sqrt{\kappa_a},\sqrt{\kappa_a},\sqrt{\kappa_b},\sqrt{\kappa_b}]$,\,  ${\bf D_3}=1/2\,\, Diag[0,0,0,0,1,1]$ and ${\bf D_4}$ is a $6\times6$ matrix with only one non-zero element, given by
\begin{equation}
 \{{\bf D_4}\}_{(2,6)}=-i\omega \frac{\sqrt{\sigma}g_{cd}t}{\sqrt{2\kappa_b}}.
\end{equation}
By inserting Eq.~(\ref{coro}) into Eq.~(\ref{vfl}) and integrating ${\bf V}^{fl}(\omega,\omega')$ using the delta function $\delta(\omega+\omega')$, the final expression of the covariance matrix of the filtered cavity output modes is given by
\begin{equation}
{\bf V}^{flt}(\Omega_a,\tau_a;\Omega_b,\tau_b)=\int^\infty_{-\infty} d\omega {\bf V}^{flt}(\omega,-\omega),
\end{equation}
where we explicitly show the dependence on the central frequencies $\Omega_j$ and inverse linewidths $\tau_j$ of the output field filter functions.


\section{Steady state entanglement, fidelity of teleportation, and two way steerability}\label{ses}
In order to study the entanglement of a traveling CV bipartite Gaussian system, composed of the filtered output optical modes $A$ (Alice) and $B$ (Bob), the covariance matrix ${\bf V}_{ab}$  of the reduced Gaussian state $\hat{\rho}_{ab}$ can be obtained by eliminating the mechanical mode, i.e, by removing the rows and columns of the covariance matrix ${\bf V}^{flt}$ corresponding to this latter mode. The resulting covariance matrix can be written in terms of $2\times2$ block matrices as
\begin{eqnarray} \label{matr}
{\bf V}_{ab}&=& \left(\begin{array}{cc}
\bf{A}& \bf{C}\\
{\bf C}^T &\bf{B}
\end{array}\right),
\end{eqnarray}
where $\bf{A}$ and $\bf{B}$ are the covariance matrices corresponding to the Alice's and Bob's subsystems respectively, whereas $\bf{C}$ describes the correlation between Alice and Bob.
The bipartite entanglement is measured by the negativity ~\cite{Vidal2002} and can be quantified using the logarithmic negativity ~\cite{Eisert2001, Adesso2004,Plenio2005} $E_N=max[0,-\mathrm{ln}(2\nu_-)]$, where $\nu_-$ is the smallest symplectic eigenvalue of partial transpose ${\bf V}_{ab}$ matrix.

When the two travelling optical output fields are entangled, they can be exploited for long-distance transfer of quantum information, e.g., for quantum teleportation of an unknown coherent state~\cite{Braunstein1998}. For long-distance applications, it is important to consider the robustness of the resulting quantum communication channel with respect to optical losses, which are unavoidable when the two fields travel a long distance in free space or down an optical fiber. Losses can be described using a beam splitter model~\cite{Barbosa2011}, with an effective transmissivity $\eta=\eta_0e^{-\alpha l/10}$, with $\alpha$ the attenuation in $dB/km$, $l$ the distance traveled by each field (assumed to be at the same distance from the generating device for simplicity), and $\eta_0$ taking into account all possible inefficiencies~\cite{Asjad2015a,Barbosa2011,Khalique2013}. Due to these losses, the filtered output covariance matrix becomes ${\bf {V}}^{loss}_{ab}=\eta {\bf {V}}_{ab}+{1\over2}(1-\eta)\bf{ I}$ with $\bf I$ the $4\times4$ identity matrix.

The fidelity for the teleportation of a Gaussian state is connected with the bipartite covariance matrix by the expression $F = 1/{\rm Det}(\bf \Gamma)$~\cite{ Barzanjeh2011, Barzanjeh2012}, with the $2\times2$ matrix ${\bf \Gamma}=2{\bf V}_{in} +{\bf B}^{loss} + {\bf Z}{\bf A}^{loss}{\bf Z} + {\bf Z}{\bf C}^{loss} + {\bf C}^{\mathsf {T}loss}{\bf Z}$, where ${\bf V}_{in}$ is the covariance matrix of the Gaussian state to be teleported, ${\bf Z}$ is the diagonal Pauli matrix, ${\bf A}^{loss}$, ${\bf B}^{loss}$and ${\bf C}^{loss}$ the matrices in Eq.~(\ref{matr}) in the presence of optical loss, while in its absence they would be the same of Eq.~(\ref{matr}). We shall always consider an input coherent state where ${\bf V}_{in}= I_2/2$, where $I_2$ is the $2\times2$ identity matrix. Moreover, the fidelity with respect to the optimal upper bound defined in Ref.~\cite{Mari2008}, obtained by optimizing over all possible local operations, is given by
 \begin{equation}\label{vm}
 F=\dfrac{1}{1+e^{-E_N}},
 \end{equation}
where $E_N$ is the logarithmic negativity of the quantum channel.

\begin{figure*}[t!]
\centering
\includegraphics[width=1\textwidth]{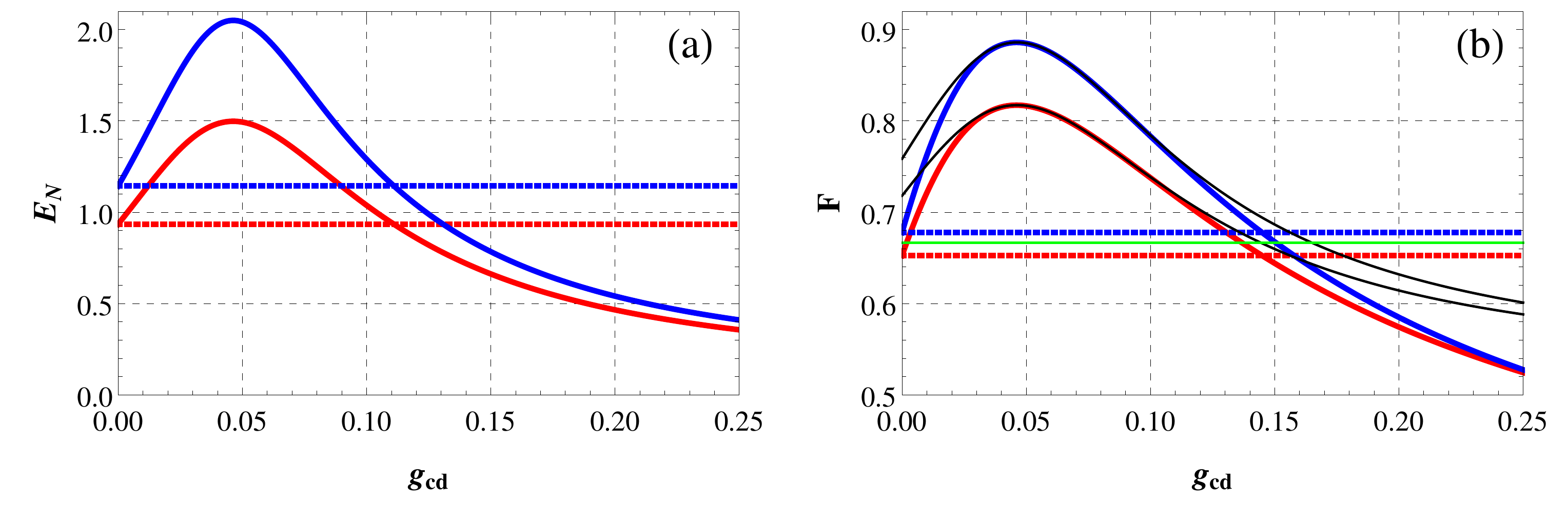}
\caption{(Color online)  Logarithmic negativity $E_N$ (a), and fidelity of teleportation of a coherent state (b), as a function of feedback gain $g_{cd}$ for fixed value of beam splitter reflectivity $r=0.56 $, both showing a maximum very close to the heating term cancelation value $g_{cd}=G_b/\sqrt{\sigma}r \omega_m$. Blue curves refer to the situation without optical loss, while red curves to that with loss, with an attenuation $\alpha=0.005 \,dB/km$ \cite{Fischer2004} in free space, distance $l=20 \,km$, and $\eta_0=0.9$. The corresponding dashed blue and red curves refer to the case without feedback $g_{cd}=0$. The green line is for the secure teleportation threshold $F_{thr}=2/3$, while the thin black curves represent the value of the upper bound defined in Eq.~(\ref{vm}). The other parameters are the same as in Fig.~(\ref{ab_r}).}\label{ab}
\end{figure*}


Let us now discuss under which conditions cold damping feedback improves the generation of CV output optical entanglement for quantum communication applications.  Cold damping feedback is typically used in the unresolved sideband regime $\kappa_b > \omega_m$ and at cavity resonance, $\Delta_b =0$, in order to optimally overdamp and cool the mechanical resonator. Here we show that, instead, the best enhancement and control of output optical entanglement is achieved in a \emph{completely different regime of cold damping}, in the resolved sideband regime $\kappa_b < \omega_m$, and where the backaction of cavity mode B and that of feedback act against each other. In fact, the best CV entanglement increase is obtained when $\Delta_b < 0$ and $g_{cd}> 0$, when mode B backaction heats and drives the mechanical resonator to instability while feedback cools and stabilizes it.

As shown in various papers \cite{Genes2009,Wang2012,Tian2012,Asjad2015a,Barzanjeh2011,Barzanjeh2012}, optimal stationary entanglement between the two output modes is achieved in the regime where one mode (here mode A) is coupled to the mechanical resonator via the beam-splitter interaction, and the other mode (mode B) via the parametric interaction, achieved when $\Delta_a=-\Delta_b=\omega_m$ and in the resolved sideband regime $\kappa_a,\kappa_b<\omega_m$, with the first interaction slightly dominant for stability conditions. If the parameters depart too much from the instability threshold, i.e., the cooling process via the beam-splitter interaction dominates too much, the entanglement degrades. We can see that stationary output mode entanglement is improved by cold damping when the latter improves the stability and cooling without modifying the coupling and detunings of mode A and B, $G_j$ and $\Delta_j$.
An intuitive idea of this fact can be obtained from the expression of the effective mechanical damping in the presence of the cavity modes backaction and feedback, that can be derived from the mechanical susceptibility \cite{Genes2009},
\begin{eqnarray}\label{gafb}
\gamma^{fb}_{m,eff}(\omega)&=&\gamma_m\left[1+\sum_{i=a,b}\dfrac{2 G^2_i\Delta_i Q_m\kappa_i}{[\kappa^2_i+(\omega-\Delta_i)^2] [\kappa^2_i+(\omega+\Delta_i)^2]}\right.\nonumber\\
&+&\left.\dfrac{(\Delta^2_b+\omega^2+\kappa^2_b)\kappa_b G_b \sqrt{\sigma}rg_{cd} Q_m}{(\kappa^2_b+(\omega-\Delta_b)^2)(\kappa^2_b+(\omega+\Delta_b)^2)}\right]
\end{eqnarray}
where $Q_m=\omega_m/\gamma_m$ is the mechanical quality factor.
When $\omega = \omega_m$, $\Delta_a=-\Delta_b=\omega_m$, and in the resolved sideband regime ($\kappa_a,\kappa_b<\omega_m$), the effective mechanical damping becomes
\begin{eqnarray}
 \gamma^{fb}_{m,eff}(\omega_m)&=&\gamma_m\left[1+\dfrac{G^2_a }{2 \kappa_a \gamma_m}-\dfrac{ G^2_b }{2\kappa_b\gamma_m}\left(1-\dfrac{\omega_m \sqrt{\sigma}rg_{cd}}{G_b}\right)\right], \nonumber \\ \label{rdgm}
 \end{eqnarray}
which shows that when $g_{cd}=G_b/\sqrt{\sigma}r\omega_m$ the heating term is canceled, implying a more stable system and a better mechanical cooling. We expect that the largest entanglement is achieved when this condition between the output beam splitter reflectivity $r$ and the feedback gain $g_{cd}$ is at least approximately satisfied.


This is confirmed in Fig.~\ref{ab_r}, where we analyze the effect of the beam splitter reflectivity at the output of mode B on the feedback performance and we show the contour-plot of the logarithmic negativity $E_N$ (Fig.~\ref{ab_r}(a)) and the teleportation fidelity for a coherent state (Fig.~\ref{ab_r}(b)) as a function of the beam splitter reflectivity $r$ and the feedback gain $g_{cd}$. Here we restrict to the case with no losses down the channel. The dashed curve corresponds just to the condition $g_{cd}~r=G_b/(\sqrt{\sigma}\omega_m)$, when cold damping feedback cancels the heating effect of the blue-detuned mode B.
We see that the optimal value of entanglement, as well as for the teleportation fidelity (practically coinciding with the upper bound of Eq. (\ref{vm})), is obtained at $g_{cd}=0.047$ and $r=0.56$. This value is quite close to one point of the dashed line $g_{cd}~r=G_b/(\sqrt{\sigma}\omega_m)$, and this suggests that cold damping feedback improves output stationary entanglement approximately when it cancels the heating effect of the blue-detuned optical mode and stabilizes the system. This effect of cold damping feedback is optimal when a beam splitter close to a $50/50$ one is used to take part of the output signal for the homodyne feedback loop, while the rest is used for the teleportation protocol.


We have also verified that when cold damping is used in a different regime one does not have the same significant enhancement of entanglement. In particular, when cold damping feedback exploits the homodyne detection of the red-detuned cooling mode $A$, feedback enforces cooling and stability as well, but one gets a smaller enhancement of entanglement. This occurs also when feedback with negative gain (i.e., anti-damping) is used, where, again, the achieved stationary entanglement is smaller. We also noticed that different feedback actions, for example the proportional feedback used for increasing the resonance frequency in~\cite{Abbott2009}, or for improving ponderomotive squeezing in~\cite{Vitali2011} is not able to provide the same entanglement's enhancement.

From now on we choose the optimal operational point $r=0.56$ and we also take the usual condition for maximizing the output entanglement in this case~\cite{Genes2009,Wang2012,Tian2012,Asjad2015a,Barzanjeh2011, Barzanjeh2012}, that is, we take for the detunings $\Delta_a=-\Delta_b=\omega_m$, and the filters' frequencies centered around the corresponding cavity mode resonance, $\Omega_a=-\Omega_b=\omega_m$.


Fig.~\ref{ab} shows the effect of the feedback gain $g_{cd}$ on the logarithmic negativity $E_N$ (Fig.~\ref{ab}(a)), and on the fidelity $F$ for the teleportation of a coherent state (Fig.~\ref{ab}(b)).
We see that cold damping feedback permits to increase the entanglement and the corresponding teleportation fidelity within an appropriate interval of values of the feedback gain, with respect to the case without feedback (dashed horizontal lines), both with (red curves) and without (blue curves) losses. Both quantities show a maximum very close to the heating term cancelation value $g_{cd}=G_b/\sqrt{\sigma}r \omega_m$, consistent with the result of Fig.~\ref{ab_r}.
We notice that the achieved maximum of $E_N$ in the presence of losses is still smaller than the largest valued achieved experimentally using parametric down conversion, i.e., by mixing two individual squeezed beams at a balanced beam
splitter ($E_N \simeq 2.3$ \cite{Eberle2013}), or at the output of a single optical parametric amplifier ($E_N \simeq  1.94$ \cite{Zhou2015}). However, as shown in \cite{ Barzanjeh2011,Barzanjeh2012,Asjad2015a,Wang2015}, one could get larger entanglement by choosing a narrower bandwidth $1/\tau_j$ ($j=a,b)$, and adjusting the corresponding couplings $G_j$, even though narrow filtered output modes are extremely difficult to prepare and to keep stable.
In the figures we consider the loss associated with free space laser communication in a clean day without turbulence, $\alpha=0.005\,dB/km$ \cite{Fischer2004}, at a distance of $l=20\,km$ and with $\eta_0=0.9$, which is equivalent to about $7\,km$ in a low noise optical fiber with attenuation $0.16\,dB/km$ \cite{Afzelius2015}. As the gain grows we pass from a regime with fidelity below the threshold for secure teleportation $F_{thr}$ to the one above, showing the advantage of closed-loop controlling the system.

\begin{figure*}[t!]
\centering
\includegraphics[width=1\textwidth]{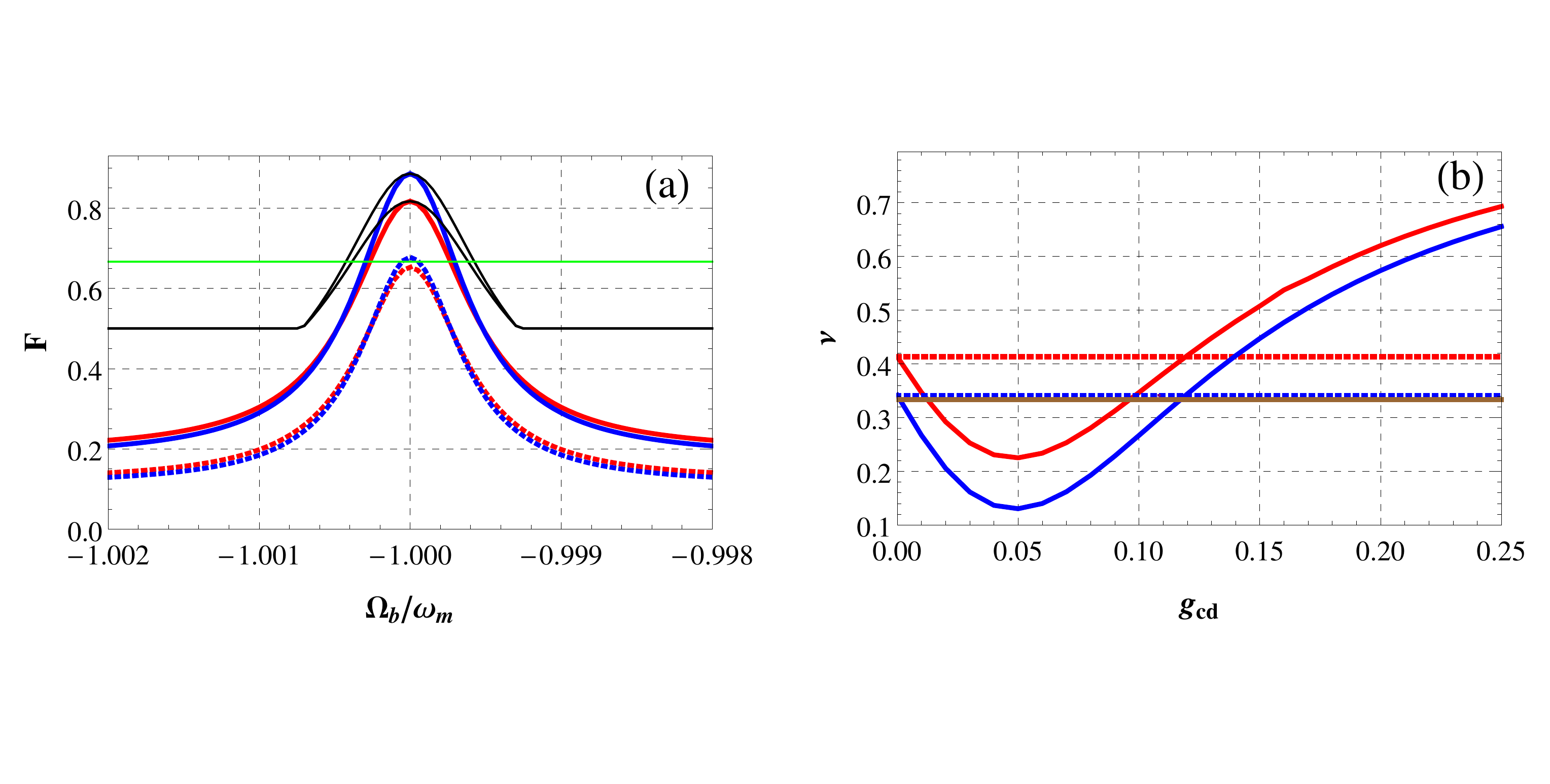}
\caption{ (Color online) (a) Fidelity of teleportation of a coherent state vs. the central frequency of one filtered output mode, with the central frequency of other output mode fixed at $\Omega_a=\omega_m$, and with $(g_{cd},r)=(0.047,0.56)$. Blue curves refer to the situation without optical loss, while red curves to that with loss. The corresponding dashed blue and red curves refer to the case without feedback $g_{cd}=0$. The green line is for the secure teleportation threshold $F_{thr}=2/3$, and black lines refer to the upper bound given in Eq.~(\ref{vm}). (b) smallest symplectic eigenvalue $\nu$ (certifying two-way steerability when $\nu<1/3$) vs. $g_{cd}$. It is evident that $\nu<1/3$ is satisfied only in a small region both with loss (red curve) and without loss (blue curve). The brown line is for $\nu=1/3$ while again the dashed lines refer to the case without feedback. All the other parameters are the same as in Fig. \ref{ab_r}.} \label{fid1}
\end{figure*}

In Fig.~\ref{fid1} (a) we study the teleportation fidelity $F$ as a function of one filter's central frequency, while the other one is fixed at the value $\Omega_a=\omega_m$. We choose the feedback gain value maximizing the fidelity in Fig.~\ref{ab}(b), and we have kept all the other parameter values as in Fig.~\ref{ab}. We see, again, in Fig.~\ref{fid1} (a) that feedback allows to surpass the threshold for quantum teleportation with losses (red solid curve), with respect to the case without cold damping feedback where the teleportation fidelity is below such a threshold (red dashed curve).

In Fig.~\ref{fid1}(b) instead we exploit the recent results of Ref.~\cite{He2015} and analyze the steering capabilities of the proposed optomechanical device. Quantum steerability~\cite{Wiseman2007,Jones2007} is stronger than entanglement and occurs when an observer (Alice) can regulate or adjust the state of another distant observer (Bob) by only local measurements performed at her side when they share an EPR entangled state~\cite{Einstein1935}. A first insight of such an EPR nonlocality was provided by the seminal paper~\cite{Drummond1990}, that characterized it in terms of violations of inferred Heisenberg uncertainty principle (see also Ref.~\cite{Giovannetti2001a}).
The steerability of a two mode CV Gaussian state $\hat{\rho}_{ab}$ $A\rightarrow B$ (Gaussian measurements at Alice site) can be quantified by
$E_{B|A}= max(0,-\mathrm{ln} \sqrt{Det {\bf\\\Upsilon}}),$
where \cite{Kogias2015,Kogias2015a}
${\bf \Upsilon}= \bf {B}- {\bf C}^T {\bf A}^{-1} {\bf C}$
is the Schur complement of ${\bf A}$ in the covariance matrix ${\bf V}_{ab}$  \cite{Wiseman2007,Jones2007,Gallego2015}. Ref.~\cite{He2015} showed that in order the sent information be really secure, i.e., one can perform secure teleportation with $F > 2/3$ employing a generic (generally non-symmetric) Gaussian bipartite state, such a state must be two-way steerable, and this is certified when $\nu<1/3$, where $\nu$  is the smaller symplectic eigenvalue of the partially transposed matrix ${\bf V}_{ab}$. The symplectic eigenvalue $\nu$ of Ref.~\protect\cite{He2015} coincides with our $2\nu_-$, due the different definition for the covariance matrix. In Fig.~\ref{fid1}(b) we see that in a narrow interval of values of the feedback gain, two-way steerability is possible even in the presence of realistic optical losses.


\section{Adding feedback by using a third mode}\label{fd3}

We are now interested in controlling the value of entanglement, i.e. to control the value of the logarithmic negativity \cite{Vidal2002, Eisert2001, Adesso2004,Plenio2005}, by introducing a further optical mode frequency (Mode C) interacting with the mechanical resonator, which is just used for the homodyne measurement and feedback. This means that in this second scheme, differently from the previous case, mode B is used only for quantum communication and not also for the homodyne measurement of its reflected part. Therefore, we should add the following equation
\begin{equation}
\dot{\hat A}_c=-(\kappa_c+i\delta_c) \hat{A}_c+ig_c \hat{A}_c \hat{q} +E_c+ \sqrt{2\kappa_c}\hat{A}^{in}_c.\label{modeC}
\end{equation}
to the above system of equations (\ref{sys1}). Then, we linearize with respect to the new steady state values, which will be formally equal to those shown above with the only difference that the index now runs as $j=(a,b,c)$. Adding the third mode and considering, of course, the new stability conditions one can show that for particular values of the renormalized coupling constant $G_c=g_c |A_{cs}|\sqrt{2}$ one could have an enhancement of the logarithmic negativity and hence of the entanglement. We now consider the feedback loop obtained by extracting a fraction
of the cavity output of Mode C, which is then processed in order to drive an appropriate actuator acting on the mechanical oscillator. We reconsider the results in Sec \ref{afb} and show that when $\delta \hat{Y}^{out}_{c}(t)=(\delta \hat{A}^{out}_{c}-\delta \hat{A}^{out \dagger}_{c})/(i\sqrt{2})$ the output quadrature of optical Mode C is detected by a balanced homodyne detector with efficiency $\sigma$, (we choose here $\sigma=1$ for simplicity) and fed back to the mechanical oscillator with some gain $g_{cd}$, we are able to control the entanglement
between the other two optical output modes. In this case, the feedback force applied to the mechanical oscillator can be written in time and frequency domain, respectively as \cite{Genes2009}
\begin{eqnarray}
f_{fb}(t)&=& \dfrac{g_{cd}}{\sqrt{2\kappa_c}}\dfrac{d}{dt}\left(\sqrt{2\kappa_c}\delta \hat Y_c(t)-\hat Y^{in}_{c}(t)\right), \nonumber \\
f_{fb}(\omega)& =&-\dfrac{i\omega g_{cd}}{\sqrt{2\kappa_c}} \left(\sqrt{2\kappa_c}\delta \hat Y_{c}(\omega)-\hat Y^{in}_{c}(\omega)\right),
\end{eqnarray}
where $g_{cd}>0$ is the feedback gain. The modified dynamics of the four-mode optomechanical system can be written in compact form in the frequency domain as
\begin{eqnarray}\label{fb3}
{\bf \hat{\mathcal{R}}}^{fb}(\omega)= -{\bf \mathcal{M}}(\omega){\bf \hat{\mathcal{N}}}^{fb}(\omega),
\end{eqnarray}
where ${\bf \hat{\mathcal{R}}}^{fb}(\omega)=[\delta \hat{q}^{fb},\delta  \hat{p}^{fb}, \delta  \hat{X}^{fb}_a,\delta  \hat{Y}^{fb}_a, \delta \hat{ X}^{fb}_b,\delta  \hat{Y}^{fb}_b, \delta \hat{ X}^{fb}_c,\delta  \hat{Y}^{fb}_c]^\mathsf{T}$ is the vector of the quadrature fluctuations and ${\bf \hat{\mathcal{N}}}^{fb}(\omega) =[0,-g_{cd}\sqrt{2\kappa_b}\hat{Y}^{in}_c(\omega)(1+ \dfrac{i\omega}{2\kappa_c}),\sqrt{2\kappa_a}  \hat{X}^{in}_a,\sqrt{2\kappa_a}  \hat{Y}^{in}_a,\sqrt{2\kappa_b} \hat{X}^{in}_b,\sqrt{2\kappa_b}  \hat{Y}^{in}_b,\sqrt{2\kappa_c} \hat{X}^{in}_c,\sqrt{2\kappa_c}  \hat{Y}^{in}_c]^\mathsf{T}$ is the corresponding vector of input noises in the presence of the optical Mode C with feedback. Moreover, ${\bf \mathcal{M}}(\omega)=(i\omega I+{\bf \mathcal{A}^{dr}})^{-1}$ where ${\bf \mathcal{A}^{dr}}$ is drift matrix in the presence of Mode $C$ with feedback and is given by
\begin{eqnarray}
{\bf \mathcal{A}^{dr}}=\left(\small \begin{array}{cccccccc}
0                     &-\omega_m &0         &0         &0        &0        &0                    &0\\
-\omega_m-g_{cd }G_c  &-\gamma_m &G_a       &0         &G_b      &0        &G_c+g_{cd}\Delta_c &g_{cd}\kappa_c       \\
0                     &0         &-\kappa_a &\Delta_a  &0        &0        &0    &0 \\
G_a                   &0         &-\Delta_a &-\kappa_a &0        &0        &0     &0\\

0                     &0         &0         &0         &-\kappa_b &\Delta_b &0    &0\\
G_b                   &0         &0         &0         &-\Delta_b &-\kappa_b &0   &0\\
0                     &0         &0         &0         &0         &0      &-\kappa_c &\Delta_c\\
G_c                   &0           &0         &0       &0         &0      &-\Delta_c  &-\kappa_c
\end{array}\right).\nonumber\\
\end{eqnarray}
\begin{figure*}[t!]
\centering
\includegraphics[width=1\textwidth]{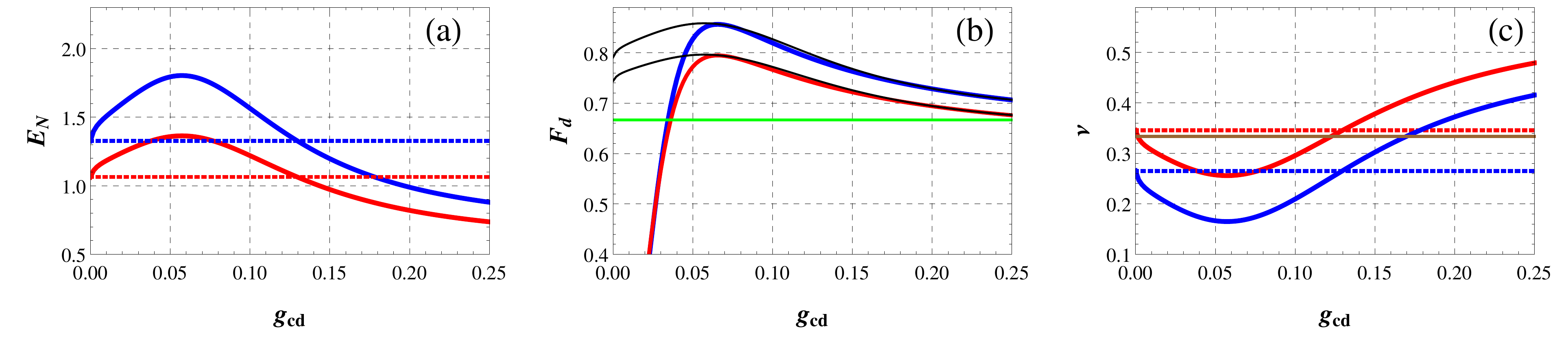}
\caption{(Color online) (a) Logarithmic negativity $E_N$ versus feedback gain $g_{cd}$ in the three-mode case. The blue curves refer to the case without loss, and the red curves to that with loss (with $\alpha=0.005 \,dB/km$ \cite{Fischer2004}, $l=20 \,km$, $\eta_0=0.9$, as in Fig. 2), and the corresponding dashed lines refer to the case without feedback. (b) Fidelity $F$ for coherent state teleportation versus $g_{cd}$, with the same color code and with the green line denoting the secure teleportation threshold $F_{thr}=2/3$. Similarly to the two optical mode case, entanglement and fidelity is maximum with very good approximation when $g_{cd}=(\kappa_c G^2_b+\kappa_b G^2_c)/(\kappa_c\omega_m G_{c})$, where the feedback cancels the heating and anti-damping effects of modes B and C. (c) smallest symplectic eigenvalue $\nu$ (certifying two-way steerability when $\nu<1/3$) vs. $g_{cd}$. It is evident that $\nu<1/3$ is satisfied only in a small region both with loss (red curve) and without loss (blue curve). The brown line is for $\nu=1/3$ while again the dashed lines refer to the case without feedback. The system parameters are $\omega_m=2\pi\times10\,\mathrm{MHz}$, $\gamma_m=1.5\times10^{-5}\,\omega_m$,  $\kappa_a=0.01\,\omega_m$, $\kappa_b=0.01\,\omega_m$, $\kappa_c=0.01\,\omega_m$, temperature $T=400\,mK$, $\omega_m \tau_b=\omega_m \tau_a=2000$, $G_a=0.065\,\omega_m$, $G_b=0.04\,\omega_m$,$G_c=0.05\,\omega_m$, $\Delta_a=\omega_m$, $\Delta_b=\Delta_c=-\omega_m$, $\Omega_{a}=-\Omega_{b}=\omega_m$. } \label{abc}
\end{figure*}
In the presence of third Mode $C$ and of feedback, the effective damping rate of the mechanical resonator can be written as (compare also with Eq. \ref{gafb})
\begin{eqnarray}\label{Gafb}
\Gamma^{fb}_{m,eff}(\omega)&=&\gamma_m\left[1+\sum_{i=a,b,c}\dfrac{2 G^2_i\Delta_i Q_m\kappa_i}{[\kappa^2_i+(\omega-\Delta_i)^2] [\kappa^2_i+(\omega+\Delta_i)^2]}\right.\nonumber\\
&+&\left.\dfrac{(\Delta^2_c+\omega^2+\kappa^2_c)\kappa_c G_c g_{cd} Q_m}{(\kappa^2_c+(\omega-\Delta_c)^2)(\kappa^2_c+(\omega+\Delta_c)^2)}\right]
\end{eqnarray}
which, in the resolved sideband regime ($\kappa_a,\kappa_b,\kappa_c<\omega_m$.), and taking the usual conditions for the detunings for maximizing the output entanglement between Mode $A$ and Mode $B$ \cite{Wang2012}, $\Delta_a=-\Delta_b=-\Delta_c=\omega_m$, becomes
\begin{eqnarray}\label{gaef}
\Gamma^{fb}_{m,eff}(\omega_m)&=&\gamma_m\left(1+\dfrac{G^2_a}{2\kappa_a\gamma_m}-\dfrac{G^2_b}{\kappa_b\gamma_m}-\dfrac{G^2_c}{2\kappa_c\gamma_m}+\dfrac{G_c\omega_m g_{cd}}{2\kappa_c\gamma_m}\right).\nonumber\\
\end{eqnarray}
It is evident that when $g_{cd}=(\kappa_cG^2_b+\kappa_b G^2_c)/\kappa_b\omega_m G_c$, the heating term corresponding to decreasing damping vanishes and feedback improves the stability of the system. We will see that, similarly to the two optical mode case, entanglement and fidelity is maximum with very good approximation close to this condition.

As in the previous Section we introduce the filters $h_j (t)$ with $j = a, b$. The output of the causal filter $h_j (t)$ can be defined by the corresponding bosonic annihilation operators as before in Eq.(\ref{filt}). The correlation matrix of the filtered quadratures modes in the presence of Mode $C$ and feedback can be written as
\begin{widetext}
\begin{eqnarray}
{\bf \mathcal{V}}^{flt}(\Omega_a,\tau_a,\Omega_b,\tau_b,)&=&\int^\infty_{-\infty}d\omega\,\{\, {\bf \mathcal{T}}(\omega)\left({\bf \mathcal{P}} {\bf \mathcal{M}}(\omega){\bf \mathcal{D}}^{fb}(\omega,-\omega){\bf \mathcal{M}}^\mathsf{T}(-\omega){\bf \mathcal{P}}+ {\bf \mathcal{D}}_1 -{\bf \mathcal{P}} {\bf \mathcal{M}}(\omega) {\bf \mathcal{D}}_2 -{\bf \mathcal{D}}_2 {\bf \mathcal{M}}^\mathsf{T}(-\omega) {\bf \mathcal{P}} \right){\bf \mathcal{T}}^\mathsf{T}(-\omega)\nonumber\\
&+& {\bf \mathcal{T}}(-\omega)\left({\bf \mathcal{P}} {\bf \mathcal{M}}(-\omega){\bf \mathcal{D}}^{fb}(-\omega,\omega) {\bf \mathcal{M}}^\mathsf{T}(\omega) {\bf \mathcal{P}}+{\bf \mathcal{D}}_1 -{\bf \mathcal{P}} {\bf \mathcal{M}}(-\omega){\bf \mathcal{D}}_2 -{\bf \mathcal{D}}_2 {\bf \mathcal{M}}^\mathsf{T}(\omega) {\bf \mathcal{P}} \right){\bf \mathcal{T}}^\mathsf{T}(\omega)\}.
\end{eqnarray}
\end{widetext}
where ${\bf\mathcal{T}}(\omega)$ is the Fourier transform of the transformation matrix $\mathcal{T}(t)$ containing the filter functions
\begin{eqnarray}
{\bf \mathcal{T}}(t)=\left(\small \begin{array}{cc}
{\bf T}(t)                & {\bf 0}_{6\times 2} \\
{\bf 0}_{2\times 6}  &\delta(t) {\bf I}_{2\times 2}  \\

\end{array}\right),
\end{eqnarray}
with ${\bf 0}$ is a $6\times 2$ null matrix and ${\bf I}$ is $2\times 2$ identity matrix,
${\bf \mathcal{P}}=Diag[1,1,\sqrt{2\kappa_a},\sqrt{2\kappa_a},\sqrt{2\kappa_b},\sqrt{2\kappa_b},1,1]$, ${\bf \mathcal{D}}_1=(1/2)\, Diag[0,0,1,1,1,1,0,0]$, ${\bf \mathcal{D}}_2=(1/\sqrt{2})\, Diag[0,0,\sqrt{\kappa_a},\sqrt{\kappa_a},\sqrt{\kappa_b},\sqrt{\kappa_b},0,0]$ and ${\bf \mathcal{D}}^{fb}(\omega,\omega')= {\bf\mathcal{Z}}+{\bf \mathcal{Z}}^{fb}_{}(\omega,\omega')$ is the diffusion matrix with
\begin{equation}
{\bf \mathcal Z}=\left( \begin{array}{cc}
{\bf d}       &{\bf 0}_{6\times 2} \\
{\bf 0}_{2\times 6} &\kappa_c {\bf I}_{2\times 2}
\end{array}\right)
\end{equation}
and ${\bf \mathcal{Z}}^{fb}$ is a $8\times 8$ matrix whose nonzero elements are $\{{\bf
 \mathcal{Z}}^{fb}\}_{22}= g^2_{cd}\kappa_c(1+\frac{i\omega}{2\kappa_c})(1+\frac{i\omega'}
{2\kappa_c})$,  $\{{\mathcal{Z}}^{fb}\}_{26}= -g_{cd}\kappa_c(1+ \frac{i\omega}{2\kappa_c})$ and
$\{{\bf\mathcal{Z}}^{fb}\}_{62}= -g_{cd}\kappa_c(1+ \frac{i\omega'}{2\kappa_c})$.


In Fig.~\ref{abc}(a) we plot the logarithmic negativity $E_N$ as a function of feedback gain $g_{cd}$ by fixing the values of the detuning $\Delta_a=-\Delta_b=-\Delta_c=\omega_m$ and central frequencies $\Omega_a=-\Omega_b=\omega_m$ of the two filters to get the maximum value of entanglement between Mode $A$ and Mode $B$ at the output of the cavity in the presence of the Mode $C$. It is evident that $E_{N}$ increases with respect to the case without feedback (blue and red dotted curves), both in the absence (blue solid curve), and in the presence (red curve) of losses. Similarly to the two optical mode case, entanglement is maximum with very good approximation when $g_{cd}=(\kappa_c G^2_b+\kappa_b G^2_c)/(\kappa_c\omega_m G_{c})$, when cold dampng cancels the heating due to the blue-detuned modes B and C. In Fig. \ref{abc}(b) we show the fidelity of the teleported initial coherent state as a function of feedback gain $g_{cd}$. The fidelity with feedback (blue and red solid curves) is higher than the one without feedback (blue and red dotted horizontal lines) as expected, and reaches the upper bound (black thin solid curves) defined in Ref.\cite{Mari2008}. In Fig. \ref{abc}(a) we plot the smallest symplectic eigenvalue as a function of $g_{cd}$, as we already did for the case with only two optical modes, in order to determine the interval of values of $g_{cd}$ in which the optomechanical scheme in the presence of feedback can be used for CV two-way steerability.

\section{Conclusion}\label{con}
Following Ref. \cite{He2015} we see that in the case of the first feedback scheme employing only two optical modes, one can achieve two-way steerability even in the presence of loss, and consequently teleportation is secure with respect to a cheating sender and an infinitely able eavesdropper, when the feedback gain is chosen in the right interval. In contrast, as shown in Fig.~\ref{fid1} (b), if we consider the same system without feedback, whether in presence or not of the same loss the resulting Gaussian bipartite state is no more two-way steerable.
Of course, as it is shown in \cite{Barzanjeh2011,Barzanjeh2012,Asjad2015a,Wang2015}, choosing a narrower bandwidth $1/\tau_j$ ($j=a,b)$ for the filters one could obtain larger entanglement, a fidelity larger than the threshold for secure teleportation, and two-way steerability also without feedback, but we remark that very narrow filtered output modes are extremely difficult to prepare and to keep stable; in our plots we considered $\tau_a=\tau_b=2000 \,\omega^{-1}_m$, which are close to the limits for a practical experimental implementation. Another possibility could be using the noiseless linear amplification as in Ref. \cite{Ulanov2015}, however not easy to implement.

When we consider instead the second feedback scheme, which employs a third optical mode, and therefore requires a more involved setup, using the same values of parameters as in the above case, two-way steerability is achieved in the absence of losses also without feedback, but it is achieved in the presence of losses down the line only employing feedback. Therefore, the security of teleportation is satisfied in a lossy channel only in this latter case, showing the relevance of using the proposed feedback scheme.

{\it Acknowledgments.\textendash}This work has been supported by the European Commission (ITN-Marie Curie project cQOM and FET-Open Project iQUOEMS),


\begin{thebibliography}{10}



\bibitem{Lloyd2000} S. Lloyd, Phys. Rev. A {\bf 62}, 022108 (2000).
\bibitem{Wiseman1993a} H. M. Wiseman and G. J. Milburn, Phys. Rev. Lett. {\bf 70}, 548 (1993).
\bibitem{Wiseman1993} H. M. Wiseman, and G. J. Milburn,  Phys. Rev. A {\bf 47}, 642 (1993).
\bibitem{Tombesi1994} P. Tombesi, and D. Vitali Phys. Rev. A {\bf 50}, 4253 (1994).
\bibitem{Tombesi1995} P. Tombesi, and D. Vitali Phys. Rev. A {\bf 51}, 4913 (1995).
\bibitem{Mancini1998} S. Mancini, D. Vitali, and P. Tombesi  Phys. Rev. Lett. {\bf 80}, 688 (1998).
\bibitem{Courty2001} J.-M. Courty, A. Heidmann, and M. Pinard, Eur. Phys. J. D \textbf{17}, 399 (2001).
\bibitem{Vitali2002} D. Vitali, S. Mancini, L. Ribichini, and P. Tombesi,
Phys. Rev. A \textbf{65} 063803 (2002); \textbf{69}, 029901(E) (2004).
\bibitem{Vitali2003} D. Vitali, S. Mancini, L. Ribichini, and P. Tombesi, J. Opt. Soc. Am. B \textbf{20}, 1054 (2003).
\bibitem{Genes2009} C. Genes, A. Mari, D. Vitali, and P. Tombesi, Phys. Rev. A {\bf 78}, 032316 (2008). {\it Adv. At. Mol.Opt.  Advances in Atomic, Molecular, and Optical Physics} {\bf 57}, 33 (2009).

\bibitem{Cohadon1999} P. F. Cohadon, A. Heidmann, and M. Pinard  Phys. Rev. Lett. {\bf 83}, 3174 (1999).
\bibitem{Metzger2004} C. H\"ohberger Metzger and K. Karrai, Nature (London) \textbf{432}, 1002 (2004).
\bibitem{Kleckner2006} D. Kleckner and D. Bouwmeester Nature {\bf 444}, 75 (2006).
\bibitem{Arcizet2006b} O. Arcizet, P.-F. Cohadon, T. Briant, M. Pinard, A. Heidmann, J.-M. Mackowski, C. Michel, L. Pinard, O. Français, and L. Rousseau,  Phys. Rev. Lett. \textbf{97}, 133601 (2006).

\bibitem{Corbitt2007} T. Corbitt, Y. Chen, E. Innerhofer, H. Müller-Ebhardt, D. Ottaway, H. Rehbein, D. Sigg, S. Whitcomb, C. Wipf, and N. Mavalvala, Phys. Rev. Lett. \textbf{98}, 150802 (2007).

\bibitem{Corbitt2007a} T. Corbitt, C. Wipf, T. Bodiya, D. Ottaway, D. Sigg, N. Smith, S. Whitcomb, and N. Mavalvala  Phys. Rev. Lett. \textbf{99}, 160801 (2007);

\bibitem{Poggio2007} M. Poggio, C. L. Degen, H. J. Mamin, and D. Rugar, Phys. Rev. Lett. \textbf{99}, 017201 (2007).

\bibitem{Wilson2015} D. J. Wilson, V. Sudhir, N. Piro, R. Schilling, A. Ghadimi, and T. J. Kippenberg, Nature \textbf{524}, 325 (2015).
\bibitem{Krause2015} A. Krause, T. D. Blasius, and O. Painter, arXiv:1506.01249 (2015).
\bibitem{Sudhir2016} V. Sudhir, D. J. Wilson, R. Schilling, H. Schütz, S. A. Fedorov, A. H. Ghadimi, A. Nunnenkamp, and T. J. Kippenberg, arXiv:1602.05942 (2016).

\bibitem{Bushev2006} P. Bushev, D. Rotter, A. Wilson, F. Dubin, C. Becher, J. Eschner, R. Blatt, V. Steixner, P. Rabl, and P. Zoller,  Phys. Rev. Lett. \textbf{96}, 043003 (2006).

\bibitem{Koch2010} M. Koch, C. Sames, A. Kubanek, M. Apel, M. Balbach, A. Ourjoumtsev, P. W. H. Pinkse, and G. Rempe, Phys. Rev. Lett. \textbf{105} 173003 (2010).


\bibitem{Gieseler2012} J. Gieseler, B. Deutsch, R. Quidant, and L. Novotny, Phys. Rev. Lett. \textbf{109}, 103603 (2012).

\bibitem{Genoni2015} M. G. Genoni, J. Zhang, J. Millen, P. F Barker and A. Serafini  New J. Phys. \textbf{17}, 073019 (2015).

\bibitem{Riste2013} D. Rist\`e, M. Dukalski, C. A. Watson, G. de Lange, M. J. Tiggelman, Ya. M. Blanter, K. W. Lehnert, R. N. Schouten, and L. DiCarlo  Nature (London) \textbf{502}, 350 (2013).
\bibitem{Vidal2002} G. Vidal and R F Werner  Phys. Rev. A {\bf 65}, 032314 (2002).
\bibitem{Eisert2001} J. Eisert PhD Thesis University of Potsdam 2001.
\bibitem{Adesso2004} G. Adesso, A. Serafini, and F. Illuminati, Phys. Rev. A {\bf 70}, 022318 (2004).
\bibitem{Plenio2005} M. Plenio, Phys. Rev. Lett. \textbf{95}, 090503 (2005); \textit{ibid.} \textbf{95} 119902 (2005).
\bibitem{Grosshans2001} F. Grosshans and P. Grangier, Phys. Rev. A {\bf 64}, 010301(R) (2001).
\bibitem{He2015} Q. He, L. Rosales-Zárate, G. Adesso, and M. D. Reid, Phys. Rev. Lett. {\bf 115}, 180502 (2015).

\bibitem{Buzek1996} V. Bu\u{z}ek and M. Hillery, Phys. Rev. A {\bf 54}, 1844 (1996).

\bibitem{Giovannetti2001a} V. Giovannetti, S. Mancini, and P. Tombesi, Europhys. Lett. {\bf 54}, 559 (2001).

\bibitem{Pinard1999} M. Pinard, Y. Hadjar, and A. Heidmann,  Eur. Phys. J. D {\bf 7}, 107 (1999).

\bibitem{Gardiner2000} C. W. Gardiner and P. Zoller, {\it Quantum Noise} (Springer, Berlin, 2000).

\bibitem{Wang2012} Y-D. Wang and A. A. Clerk, Phys. Rev. Lett. \textbf{108}, 153603
(2012).

\bibitem{Tian2012} L. Tian, Phys. Rev. Lett. \textbf{108}, 153604 (2012).

\bibitem{Kuzyk2013} M. C. Kuzyk, S. J. van Enk, and H. Wang Phys. Rev. A \textbf{88} 062341 (2013).
\bibitem{Asjad2015a} M. Asjad, S. Zippilli, P. Tombesi, and D. Vitali, Phys. Scr. {\bf 90} 074055 (2015).



\bibitem{Barzanjeh2011} Sh. Barzanjeh, D. Vitali, P. Tombesi, and G. J. Milburn, Phys. Rev. A {\bf 84}, 042342 (2011).

\bibitem{Barzanjeh2012} Sh. Barzanjeh, M. Abdi, G. J. Milburn, P. Tombesi, and D. Vitali, Phys. Rev. Lett. {\bf 109}, 130503 (2012).

\bibitem{Gopal2002} M. Gopal {\it Control Systems: Principles and Design}, (Tata McGraw-Hill Education 2002).

\bibitem{Braunstein1998} S. L. Braunstein and H. J. Kimble, Phys. Rev. Lett. {\bf 80}, 869 (1998).

\bibitem{Barbosa2011}
F. A. S. Barbosa, A. J. de Faria, A. S. Coelho, K. N. Cassemiro, A. S. Villar, P. Nussenzveig, and M. Martinelli, Phys. Rev. A {\bf 84}, 052330 (2011).

\bibitem{Khalique2013} A. Khalique, W. Tittel, and B. C. Sanders, Phys. Rev. A {\bf 88}, 022336 (2013).
%

\bibitem{Mari2008} A. Mari, and D. Vitali, Phys. Rev. A {\bf 78}, 062340 (2008).

\bibitem{Abbott2009} B. Abbott {\it et al.}, New J. Phys. \textbf{11} 073032 (2009).

\bibitem{Vitali2011} D. Vitali and P. Tombesi, C. R. Physique., {\bf 12}, 848 (2011).




\bibitem{Eberle2013} T. Eberle, V. H\"andchen, and R. Schnabel, Opt. Express \textbf{21}, 11546 (2013).

\bibitem{Zhou2015} Y. Zhou, X. Jia, F. Li, C. Xie, and K. Peng, Opt. Express \textbf{23}, 4952 (2015).



\bibitem{Wang2015} Y-D. Wang, S. Chesi, and A. A. Clerk, Phys. Rev. A {\bf 91}, 013807 (2015).



\bibitem{Fischer2004} K. W. Fischer, M. R. Witiw, J A. Baars, and T. R. Oke, Bull. Amer. Meteor. Soc. {\bf 85}, 725 (2004).

\bibitem{Afzelius2015} M. Afzelius, N. Gisin, and H de Riedmatten, Phys. Today {\bf 68} (n. 12), 42 (2015).
\bibitem{Wiseman2007} H.M. Wiseman, S.J. Jones, A.C. Doherty,  Phys. Rev. Lett. {\bf 98}, 140402 (2007).

\bibitem{Jones2007} S. J. Jones, H. M. Wiseman, and A. C. Doherty, Phys. Rev. A {\bf 76}, 052116 (2007).
\bibitem{Einstein1935} A. Einstein, B. Podolsky, and N. Rosen, Phys. Rev. {\bf 47}, 777 (1935).
\bibitem{Drummond1990}
P. D. Drummond and M. D. Reid, Phys. Rev. A {\bf 41}, 3930 (1990).
\bibitem{Kogias2015} A. Kogias, R. Lee, S. Ragy, and G. Adesso, Phys. Rev. Lett. {\bf 114}, 060403 (2015).
\bibitem{Kogias2015a}  A. Kogias and  G. Adesso, J. Opt. Soc. Am. B {\bf 32} 4 (2015).
\bibitem{Gallego2015} R. Gallego and L. Aolita, Phys. Rev. X {\bf 5}, 041008 (2015).
\bibitem{Ulanov2015} A. E. Ulanov, I. A. Fedorov, A. A. Pushkina, Y. V. Kurochkin, T. C. Ralph, A. I. Lvovsky, Nature Photonics  {\bf 9}, 764 (2015).


\end{thebibliography}
\end{document}